\documentclass[11pt]{article}

\begin{document}
\def\a{\alpha}\def\b{\beta}\def\g{\gamma}\def\d{\delta}\def\e{\epsilon}
\def\k{\kappa}\def\l{\lambda}\def\L{\Lambda}\def\s{\sigma}\def\S{\Sigma}
\def\Th{\Theta}\def\th{\theta}\def\om{\omega}\def\Om{\Omega}\def\G{\Gamma}
\def\y{\vartheta}\def\m{\mu}\def\n{\nu}
\def\ws{worldsheet}
\def\susy{supersymmetry}
\def\ts{target superspace}
\def\ks{$\k$--symmetry}
\newcommand{\plabel}{\label}
\renewcommand\baselinestretch{1.5}
\newcommand{\nn}{\nonumber\\}\newcommand{\p}[1]{(\ref{#1})}
\renewcommand{\thefootnote}{\arabic{footnote}}

\thispagestyle{empty}

\begin{flushright}
hep-th/0102109
\end{flushright}

\vspace*{1cm}
\begin{center}
{\Large \bf  T-duality in the string theory effective action
\\ with a string source
}
\vspace*{2cm}
\\
Alexei Nurmagambetov
\footnote{e-mails:
alexei.nurmagambetov@pd.infn.it, ajn@kipt.kharkov.ua}\\
\vspace{0.5cm}
{\it Institute for Theoretical Physics}\\
{\it NSC ``Kharkov Institute of Physics
and Technology"}\\
{\it 61108, Kharkov, Ukraine}
\footnote{Permanent position.}
\\ 
{\it and}\\
{\it Dipartimento di Fisica ``Galileo Galilei"}\\
{\it Universit\`a di Padova}\\
{\it Via F. Marzolo, 8}\\
{\it 35131, Padova, Italia}
\footnote{Postdoctoral fellow.}
\\ \bigskip

\vspace{1.5cm}
{\bf Abstract}
\end{center}

\noindent
We consider the T-duality transformations of the low-energy quantum
string theory effective action in the presence of classical fundamental 
string source and demonstrate explicitly that T-duality still holds.


\vspace{2.8cm}


\vspace{0.8cm}

\newpage

\section{Introduction}

Target-Space duality (or T-duality) is the one of the dualities
set of superstring theory which can be viewed already at the
perturbative level. As it has been pointed out in papers by
Kikkawa and Yamasaki \cite{ky} and Sakai and Senda \cite{ss} 
it arises the formal symmetry of quantum effective potential of 
free toroidally compactified string theory
under interchange of a
torus radii with their inverse. In its turn this phenomenon occurs
in the firstly quantized string theory by virtue of symmetry of
the ${\sf mass}^2$ operator under $R \leftrightarrow 1/R$
interchange supplemented by interchange of Kaluza-Klein and
winding modes arising under compactification.
The next important step was made in paper by Nair, Shapere,
Strominger and Wilczek \cite{dssw} where this symmetry was
justified in the interacting case as well, and therefore received the
status of exact symmetry of firstly quantized string theory.
However only after the seminal papers by Buscher \cite{buscher} it
became clear that at the level of classical string dynamics this
symmetry is realized as impossibility to differentiate (special
kind of or so-called dual) backgrounds with (one or many)
isometry directions in which string propagates. Connection
between such kind of backgrounds coming from the target-space
metric, the 2-form antisymmetric tensor field and the dilaton is
given by the set of Buscher's formulae
\footnote{The analog of these formulas for the target-space metric was derived
independently by Hitchin, Karlhede, Lindstr{\"o}m and Ro{\v{c}}ek
in \cite{hklr} addressed to investigation of $d$-dimensional
non-linear sigma-model.}
\cite{buscher}.

Consideration of the long-wave limit of superstring theory
corresponding to supergravity action also leads to the
establishing the symmetry between dimensionally reduced dual
theories \cite{hw}--\cite{bho} and allows to exhibit more clearly
the space-time interpretation of the duality transformations as
ones acting only on the matter fields and leaving the
gravitational sector unchanged. Also, the target-space geometry
adapted to the T-duality transformations was constructed in
\cite{borlaf}.

What concerns supersymmetric string theory, the T-duality
transformations were intensively investigated in the NSR
(worldsheet SUSY) formulation (see, for instance
\cite{agb}--\cite{abb} and Refs. therein), although
the absence of direct connection between worldsheet and
target-space supersymmetries, which can be established with
application of the CFT technique \cite{bd}, hampered the problem of
getting the T-duality rules for the space-time fermions (see, however, 
\cite{hassan1}). This
drawback naturally overcomes in the manifestly target-space
supersymmetric formulation of the GS superstring, however the
progress in the construction of T-duality rules in the GS
formulation
\cite{clps}--\cite{kr} was achieved only recently.

Therefore, superstring and supergravity encourage the
exact symmetry under the T-duality transformations. Moreover, as it has been
pointed out in literature (see e.g. \cite{t}) the sum of supergravity action 
and fundamental
superstring source action should be T-duality covariant. The goal
of this communication is to push forward this observation and
to establish the invariance of dual
theories under the T-duality transformations for the toy model of
closed bosonic string being the source for gravity and 2-form
antisymmetric tensor field. Because our consideration is pure
bosonic we will to refer in what follows this bulk configuration
as low energy quantum string theory effective action \cite{ft0} or effective
action for shortness. It is clear that such a configuration
is the part of the bosonic sector of supergravity being the superstring theory 
effective action. Another point of our consideration consists in the  
restriction of our analysis
to the pure classical frames. It means that we shall not
consider the dilaton appearance in the string action and, consequently,
in the set of the T-duality rules, because it is 
well-known that the dilaton shift appearing in the Buscher's formulae can be
described correctly only in the frame of quantum approach
\cite{abl}, \cite{jrst} (and Refs. therein). However, since supergravity
contains the information about quantum corrections of the worldsheet, we have 
to take into account the dilaton shift as far as the effective action will be
considered.

To reach the goal we analyse two different approaches which lead to the result.
The first approach is the standard one and it demonstrates the role of 
worldvolume duality in establishing the T-duality invariance in the bulk. 
The second approach we propose is based on the gauging of the translational 
invariance along isometry direction simultaneously in the bulk and in the 
source parts of the effective action with source. This forces us, in 
particular, to introduce new additional target-space field instead of 
worldvolume one as it is made in the first case.

The paper is organized as follows. In Section 2 we give a brief
review of derivation of the Buscher's formulae in the standard
sigma-model-like \cite{buscher} and gauged sigma-model-like 
\cite{hs0}--\cite{ps} approaches. Section
3 devoted to the recalling the T-duality rules in 
Kaluza-Klein picture \cite{bek,bko}, which simplify calculations
(see, for instance, \cite{bm} for non-Abelian case, and Ref. \cite{borlaf}, 
where
such kind of transformations arise in the geometry of a target-space adapted to
the presence of isometries)
and in Section 4 we outline the application of
these rules to the
gauged-sigma-model. After these preliminaries in Section 5 we
demonstrate the invariance between dual effective theories in 
presence of the fundamental string source. Discussion of the
obtained results and concluding remarks are
collected in the last section.

\section{T-duality in bosonic string theory}

Consider Polyakov's action for bosonic (closed) string \cite{polyakov}
\begin{equation}\plabel{SA}
S=\int\,d^2 \xi \sqrt{-\g}\g^{ij}\partial_i X^{\underline m}
\partial_j X^{\underline n}g_{\underline{mn}}(X).
\end{equation}
To reach T-dual action, let's suppose that we have an isometry direction,
say $X^{\underline{0}}$ 
\footnote{This notation rises to the notations of Ref. \cite{buscher} and 
we would like to keep it suggesting that there will not be a confusion between
isometry and time-like coordinates.}
, and evidently
\begin{equation}\plabel{SA1}
S=\int\,d^2 \xi \sqrt{-\g}\g^{ij}(\partial_i X^{\underline 0}
\partial_j X^{\underline 0}g_{\underline{00}}+
\partial_i X^{\underline 0}
\partial_j X^{\underline{\tilde m} }g_{\underline{0}\underline{\tilde m}}
+\partial_i X^{\underline{\tilde m} } \partial_j X^{\underline 0}
g_{\underline{\tilde m}\underline{0}}+
\partial_i X^{\underline{\tilde m} }\partial_j X^{\underline{\tilde n} }
g_{\underline{\tilde m}\underline{\tilde n}}).
\end{equation}
After that one can write down \p{SA1} in the first order form 
\cite{hklr}, \cite{ft} as
$$
S=\int\,d^2 \xi[
\sqrt{-\g}\g^{ij}(C_i C_j g_{\underline{00}}+ C_i \partial_j
X^{\underline{\tilde m}}g_{\underline{0}\underline{\tilde m}}
+\partial_i X^{\underline{\tilde m} } C_j g_{\underline{\tilde
m}\underline{0}}+
\partial_i X^{\underline{\tilde m} }\partial_j X^{\underline{\tilde n} }
g_{\underline{\tilde m}\underline{\tilde n}})
$$
\begin{eqnarray}\plabel{SA2}
-2\e^{ij}C_i \partial_j {\hat X}^{\underline{0}}].
\end{eqnarray}
Eq. of motion for ${\hat X}^{\underline{0}}$ field enforces the curl free
condition for worldvolume variables $C_i$, and, therefore, locally
\begin{equation}\plabel{A}
C_i=\partial_i {\tilde X}^{\underline{(0)}}.
\end{equation}
Strictly speaking ${\tilde X}^{\underline{(0)}}$ is different from
$X^{\underline{0}}$. However, we can always sew their by virtue of the
invariance of \p{SA1} under $X^{\underline{0}} \rightarrow X^{\underline{0}}
+const $, so it can be supposed that $C_i=\partial_i X^{\underline{0}}$
and the action \p{SA1} is recovered.

Another story begins if one integrates out the new variables $C_i$. To see
it, it is convenient to record the action \p{SA1} in the form of
\begin{equation}\plabel{SA1F}
S=\int_{{\cal M}^2}\, dX^{\underline m}\wedge \ast dX^{\underline n}
g_{\underline{mn}}
\end{equation}
and to make the same procedure as in \p{SA2}. Therefore
\begin{equation}\plabel{SA2F}
S=\int_{{\cal M}^2}\, C\wedge \ast C g_{\underline{00}}+
C\wedge \ast dX^{\underline{\tilde n}}g_{\underline{0}\underline{\tilde
n}}+dX^{\underline{\tilde m}}\wedge \ast C g_{\underline{\tilde m}
\underline{0}}+dX^{\underline{\tilde m}}\wedge \ast
dX^{\underline{\tilde n}}g_{\underline{\tilde m}\underline{\tilde n}}
-2C\wedge d{\hat X}^{\underline{0}}.
\end{equation}
Integrating out the $C$ field we derive
$$
{\overleftarrow{ \d}{\cal L}\over \d C}=\ast C
g_{\underline{00}}+\ast dX^{\underline{\tilde
n}}g_{\underline{0}\underline{\tilde n}}- d{\hat
X}^{\underline{0}}=0,
$$
\begin{equation}\plabel{A1}
C=-{1\over g_{\underline{00}}}(i_0 g-\ast d{\hat X}^{\underline{0}}),
\qquad
\ast C=-{1\over g_{\underline{00}}}(\ast i_0 g-d{\hat
X}^{\underline{0}})
\end{equation}
and
\begin{equation}\plabel{A2}
S=\int_{{\cal M}^2}\, dX^{\underline{\tilde m}}\wedge \ast
dX^{\underline{\tilde n}}g_{\underline{\tilde m}\underline{\tilde n}}
-{1\over g_{\underline{00}}}i_0 g\wedge \ast i_0 g+{1\over
g_{\underline{00}}} d{\hat X}^{\underline 0}\wedge \ast d{\hat
X}^{\underline 0} +{2\over g_{\underline{00}}}i_0 g \wedge d{\hat
X}^{\underline 0}.
\end{equation}
Important point is that the action \p{A2} written in the dual with respect to
the $C$ field variables $d{\hat X}^{\underline 0}$ still allows for eq. \p{A},
however it becomes ``Bianchi identity'' for the solution to the equation 
of motion for $C$ following from the original action \cite{hklr} 
(see also Refs. \cite{duff} and \cite{hl}).

It is easy to see that this action is equivalent to the string action
propagating in dual background, which is parameterized by the coordinates
${\tilde X}^{\underline m}=(X^{\underline{\tilde m}},{\hat X}^{\underline
0})$, with the NS 2-form potential ${\tilde B}^{(2)}$, i.e.
\begin{equation}\plabel{A2D}
S=\int_{{\cal M}^2}\, d{\tilde X}^{\underline m}\wedge \ast
d{\tilde X}^{\underline n}{\tilde g}_{\underline{mn}}+2{\tilde B}^{(2)},
\end{equation}
where \cite{buscher}
$$
{\tilde g}_{\underline{\tilde m}\underline{\tilde n}}=
g_{\underline{\tilde m}\underline{\tilde n}}-{1\over g_{\underline{00}}}
g_{\underline{0}\underline{\tilde m}}g_{\underline{0}\underline{\tilde
n}},\qquad {\tilde g}_{\underline{00}}={1\over g_{\underline{00}}},
$$
\begin{equation}\plabel{TR}
{\tilde B}^{(2)}_{\underline{0}\underline{m}}=
-{\tilde B}^{(2)}_{\underline{m}\underline{0}}=
{g_{\underline{0}\underline{\tilde m}}\over g_{\underline{00}}}, \qquad
{\tilde B}^{(2)}_{\underline{\tilde m}\underline{\tilde n}}=0.
\end{equation}

Therefore, target-space duality switches the pure gravitational background to
the dual background with gravity and the NS 2-form gauge field. However, if we 
start from a string in the background of the NS two-form field $B^{(2)}$ 
\begin{equation}\plabel{AB1}
S=\int_{{\cal M}^2}\, dX^{\underline m}\wedge \ast dX^{\underline n}
g_{\underline{mn}}+2B^{(2)}
\end{equation}
and apply the scheme noticed above again, we arrive at
$$
S=\int_{{\cal M}^2}\, dX^{\underline {\tilde m}}\wedge \ast dX^{\underline
{\tilde n}} g_{\underline{\tilde m}\underline{\tilde n}}
-{1\over g_{\underline{00}}}i_0 g\wedge \ast i_0 g
+{1\over g_{\underline{00}}}(d{\hat X}+i_0 B)\wedge \ast (d{\hat X}+i_0 B)
$$
\begin{equation}\plabel{AB2}
+{2\over g_{\underline{00}}}i_0 g\wedge (d{\hat X}+i_0 B)
+dX^{\underline {\tilde m}}\wedge dX^{\underline {\tilde n}}
B^{(2)}_{\underline{\tilde n}\underline{\tilde m}},
\end{equation}
which is the action for the string in dual background
\begin{equation}\plabel{AB2D}
S=\int_{{\cal M}^2}\, d{\tilde X}^{\underline m}\wedge \ast
d{\tilde X}^{\underline n}{\tilde g}_{\underline{mn}}+2{\tilde B}^{(2)}
\end{equation}
with
$$
{\tilde g}_{\underline{\tilde m}\underline{\tilde n}}=
g_{\underline{\tilde m}\underline{\tilde n}}-{1\over g_{\underline{00}}}
(g_{\underline{0}\underline{\tilde m}}g_{\underline{0}\underline{\tilde
n}}-i_0 B^{(2)}_{\underline{\tilde m}} i_0 B^{(2)}_{\underline{\tilde n}})
,\qquad {\tilde g}_{\underline{00}}={1\over g_{\underline{00}}},
$$
\begin{equation}\plabel{TRB}
{\tilde B}^{(2)(-)}=B^{(2)(-)}+{i_0 g\over g_{\underline{00}}}\wedge
i_0 B^{(2)},\ \ \ \
i_0 {\tilde B}^{(2)}={i_0 g\over g_{\underline{00}}},\ \ \ \
i_0 B^{(2)}={i_0 {\tilde g}\over {\tilde g}_{\underline{00}}},
\end{equation}
where $i_0 g\equiv dX^{\underline{\tilde
m}}g_{\underline{0}\underline{\tilde m}}$, $i_0 B\equiv
dX^{\underline{\tilde m}}B_{\underline{0}\underline{\tilde m}}$ and
$B^{(2)(-)}\equiv {1\over 2} dX^{\underline{\tilde m}}\wedge
dX^{\underline{\tilde n}}B^{(2)}_{\underline{\tilde n}\underline{\tilde
m}}$.

Another way to get the same result is to consider the isometry
gauging procedure \cite{hs0}--\cite{ps}. To recall, remind that as we have 
pointed above
the action \p{SA2} possesses the invariance under the constant
shift in the $X^{\underline{0}}$ direction, i.e. under
$X^{\underline{0}} \rightarrow X^{\underline{0}}~+const$. Let us
now to require the invariance of the action under the local shift
with some arbitrary function $f(X^{\underline{\tilde m}})$. In
this case the invariance is spoiled by the terms constructed out
the differentials of new function $f(X^{\underline{\tilde m}})$.
To compensate this contribution it is necessary to extend usual
derivative with vector field as
\begin{equation}\plabel{DX}
DX^{\underline m}=dX^{\underline m}+Ck^{\underline m},
\end{equation}
where $k^{\underline m}$ is the Killing vector in the isometry
direction. Then, in the so-called adapted frame where
$k^{\underline m}=\d^{\underline m}_{\underline 0}$ the covariant
derivative $DX^{\underline 0}$ is inert under $\d X^{\underline
0}=f$ supported by $\d C=-d \e$ with local parameter 
$\e(X^{\underline{\tilde m}})$.
Therefore, in this picture the action \p{SA2F} becomes
\begin{equation}\plabel{SD}
S=\int_{{\cal M}^2}\, DX^{\underline m}\wedge \ast DX^{\underline
n}g_{\underline {mn}}-2C\wedge d\hat{X}^{\underline 0}.
\end{equation}
Equation of motion for the $\hat{X}^{\underline 0}$ gives the curl
free condition for the field $C$ and therefore, locally we can
solve it as $C=d\tilde{X}^{\underline 0}$. By virtue of the gauge
symmetry for the vector field $C$ we can fix the gauge $C=0$ or
${X}^{\underline 0}=0$. In both cases the action \p{SD} reduces to
the action \p{SA1F} (either in terms of $(X^{\underline 0},
X^{\underline{\tilde m}})$ coordinates or $(\tilde{X}^{\underline
0}, X^{\underline{\tilde m}})$ ones). In its turn the integration
over the $C$ field in the adapted coordinate frame leads after
imposing the gauge fixing ${X}^{\underline 0}=0$ to the Buscher's
rules \p{TR}.

\section{T-duality in Kaluza-Klein picture}

As it has been outlined in Introduction,
at the level of firstly quantized string theory T-duality deals
with compactification and interchange of Kaluza-Klein (KK) and winding
modes. Let's analyse this picture from the point of view of T-duality
rules \p{TRB}. Namely, the first rule for $\tilde{g}_{\underline{\tilde m}
\underline{\tilde n}}$ can be represented as
$$
\tilde{g}_{\underline{\tilde m}\underline{\tilde n}}
-{1\over g_{\underline{00}}}i_0 B^{(2)}_{\underline{\tilde m}}
i_0 B^{(2)}_{\underline{\tilde n}}=g_{\underline{\tilde m}\underline{\tilde n}}
-{1\over g_{\underline{00}}}g_{\underline{0}\underline{\tilde m}}
g_{\underline{0}\underline{\tilde n}} \qquad  \Longrightarrow
$$
\begin{equation}\plabel{TRg1}
\tilde{g}_{\underline{\tilde m}\underline{\tilde n}}
-\tilde{g}_{\underline{00}}{i_0 \tilde{g}_{\underline{\tilde m}}\over
\tilde{g}_{\underline{00}}}
{i_0 \tilde{g}_{\underline{\tilde n}}\over
\tilde{g}_{\underline{00}}}=g_{\underline{\tilde m}\underline{\tilde n}}
-g_{\underline{00}} {i_0 g_{\underline{\tilde m}}\over
g_{\underline{00}}}
{i_0 g_{\underline{\tilde n}}\over
g_{\underline{00}}},
\end{equation}
where we have used the rules for other fields. The expressions in both sides
of \p{TRg1} are nothing but the elements $\hat{g}_{\underline{\tilde m}
\underline{\tilde n}}$ and $\hat{\tilde g}_{\underline{\tilde m}
\underline{\tilde n}}$ of the metric tensor matrix decomposition
$$
\overline{g}_{\underline{\tilde m}\underline{\tilde n}}=
\left(\begin{array}{lr}
\hat{g}_{\underline{\tilde m}
\underline{\tilde n}}:=
g_{\underline{\tilde m}\underline{\tilde n}}-{1\over
g_{\underline{0}\underline{0}}}i_0 g_{\underline{\tilde m}} i_0
g_{\underline{\tilde n}}
&
g_{\underline{0}\underline{0}} {i_0
g_{\underline{\tilde m}}\over g_{\underline{0}\underline{0}}}
\\
g_{\underline{0}\underline{0}} {i_0
g_{\underline{\tilde n}}\over g_{\underline{0}\underline{0}}}
& g_{\underline{0}\underline{0}}
\end{array} \right),
$$
$$
\overline{\tilde g}_{\underline{\tilde m}\underline{\tilde n}}=
\left(\begin{array}{lr}
\hat{\tilde g}_{\underline{\tilde m}
\underline{\tilde n}}:=
\tilde{g}_{\underline{\tilde m}\underline{\tilde n}}-{1\over
{\tilde g}_{\underline{0}\underline{0}}}i_0{\tilde g}_{\underline{\tilde m}}
i_0{\tilde g}_{\underline{\tilde n}}
&
\tilde{g}_{\underline{0}\underline{0}} {i_0
{\tilde g}_{\underline{\tilde m}}\over {\tilde g}_{\underline{0}\underline{0}}}
\\
{\tilde g}_{\underline{0}\underline{0}} {i_0
\tilde{g}_{\underline{\tilde n}}\over {\tilde g}_{\underline{0}\underline{0}}}
& {\tilde g}_{\underline{0}\underline{0}}
\end{array} \right),
$$
respecting to the following KK backgrounds
\footnote{Strictly speaking, these are effective ``KK backgrounds'' since their
form is derived from $D$-dimensional line elements 
$d{\overline s}^2={\overline g}_{\underline{m}\underline{n}}dX^{\underline m}
\otimes dX^{\underline n}$ and
$d{\overline{\tilde s}}^2={\overline {\tilde g}}_{\underline{m}\underline{n}}
dX^{\underline m}\otimes dX^{\underline n}$ by extracting $X^{\underline 0}$
coordinate and rearranging the terms in T-duality invariant combinations.}
\begin{equation}\plabel{intdir}
d\overline{s}^2=\hat{g}_{\underline{\tilde m}\underline{\tilde n}}
dX^{\underline{\tilde m}}\otimes dX^{\underline{\tilde n}}+g_{\underline{00}}
(dX^{\underline{0}}+A)\otimes (dX^{\underline{0}}+A)
\end{equation}
and
\begin{equation}\plabel{intdua}
d\overline{\tilde s}^2=
\hat{\tilde g}_{\underline{\tilde m}\underline{\tilde n}}
dX^{\underline{\tilde m}}\otimes dX^{\underline{\tilde n}}
+\tilde{g}_{\underline{00}}
(dX^{\underline{0}}+\tilde{A})\otimes (dX^{\underline{0}}+\tilde{A})
\end{equation}
with $A=i_0 g/g_{\underline{00}}$ and $\tilde{A}=i_0 \tilde{g}/
\tilde{g}_{\underline{00}}$. Therefore, in the KK picture T-duality rules
look like \cite{bko}, \cite{clps}
$$
\hat{g}_{\underline{\tilde m}\underline{\tilde n}}=
\hat{\tilde g}_{\underline{\tilde m}\underline{\tilde n}},\qquad
g_{\underline{00}}={1\over {\tilde g}_{\underline{00}}},
$$
$$
{\tilde B}^{(2)-}=B^{(2)-}+A\wedge i_0 B^{(2)},
$$
\begin{equation}\plabel{TRBKK}
A=i_0 \tilde{B}^{(2)},\qquad \tilde{A}=i_0 B^{(2)}
\end{equation}
and read off the interchange of the KK and winding modes. Apparently that
$g_{\underline{00}}$ determines the length of the internal circle, and
therefore there is also $R \leftrightarrow 1/R$.

\section{Gauged sigma-model representation and T-duality}

Let's turn the attention to the gauged sigma-model representation of string
action (eq. \p{SD}). One of the powerful points of such a representation
consists in the possibility to arrive at the result of integration over
$C$ field (eqs. \p{A2} and \p{AB2}) effectively without performing the
integration. To this end let us regard the field $C$ (or ${\tilde C}$ in dual
picture) as {\it additional} worldvolume field (as it should be),
and starting from the dual string action
\begin{equation}\plabel{SDd}
S=\int_{{\cal M}^2}\, DX^{\underline m}\wedge \ast DX^{\underline n}
{\tilde g}_{\underline{m}\underline{n}}+2{\tilde B}^{(2)}(D)
-2{\tilde C}\wedge d{\hat X}^{\underline 0}
\end{equation}
rewrite it as
$$
S=\int_{{\cal M}^2}\, [dX^{\underline{\tilde m}}\wedge \ast
dX^{\underline{\tilde n}}
\hat{\tilde g}_{\underline{\tilde m}\underline{\tilde n}}
+{\tilde g}_{\underline{00}}(dX^{\underline 0}+{\tilde C}
+{i_0 {\tilde g}\over {\tilde g}_{\underline{00}}})\wedge \ast
(dX^{\underline 0}+{\tilde C}
+{i_0 {\tilde g}\over {\tilde g}_{\underline{00}}})
$$
\begin{equation}\plabel{SDd1}
+2{\tilde B}^{(2)-}(d)+2i_0 {\tilde B}^{(2)}\wedge (dX^{\underline 0}+
{\tilde C})-2{\tilde C}\wedge d{\hat X}^{\underline 0}].
\end{equation}
Now solving for the equation for ${\hat X}^{\underline 0}$
$$
{\tilde C}=dy,~~~(can~~be~~gauged~~out)
$$
and applying the T-duality rules \p{TRBKK} we are left with expression \p{AB2}
(in the coinciding original and dual coordinates basis), which was obtained 
from the original string action \p{SA}--\p{SA2} by use of 
equation of motion for the $C$ field, i.e.
\begin{equation}\plabel{conn}
S=\int_{{\cal M}^2}\,
{\cal L}^{dual~gaug.s.mod.}_{\vert_{KK~ T-duality~ rules}}
=\int_{{\cal M}^2}\, {\cal L}^{orig.}
_{\vert_{{\d {\cal L}^{orig.}\over \d C}=0}}.
\end{equation}
Actually, this statement is trivial, because it says about equivalence of
the original theory written in terms of dual variables (after resolving for the
equation of motion for $C$ field) to its dual theory written in
terms of the same variables (after application of T-duality rules).  However,
this notion becomes crucial in the consideration of T-duality in the
effective action constructed out the part of the bosonic sector of supergravity
with the fundamental string as a source.

\section{Effective action with a string source and
T-duality}

After preliminaries made in the sections before we are in position to
consider the model describing the effective action with
string as a source for gravity and 2-form antisymmetric tensor field.
This action has the following form \cite{dkl}
$$
S=\int_{{\cal M}^D}\, e^{-2\phi}\left[ {1\over (D-2)!}R^{{\underline
a}_1{\underline a}_2}E^{{\underline a}_3}\dots E^{{\underline
a}_D}\e_{{\underline a}_1\dots {\underline a}_D} 
+{1\over 2}d\phi \ast d\phi
+{1\over
2}H^{(3)}\ast H^{(3)} \right]
$$
\begin{equation}\plabel{Seff}
-{1\over 2}\int_{{\cal M}^D}\,d^D x \sqrt{-g}{\d^D (x-X(\xi))\over
\sqrt{-g}}
\int_{{\cal M}^2} d^2 \xi
[\sqrt{-\g}\g^{ij}\partial_i X^{\underline m}
\partial_j X^{\underline n}g_{\underline{mn}}(x)
+\e^{ij}\partial_i X^{\underline m}\partial_j X^{\underline n}
B^{(2)}_{\underline{n}\underline{m}}],
\end{equation}
where $H^{(3)}=dB^{(2)}$ and the wedge product is assumed. The
variation of this action over $g_{\underline{m}\underline{n}}$
and $B^{(2)}_{\underline{m}\underline{n}}$ leads to the Einstein
equation for gravity with the dilaton and the NS 2-form gauge field in the 
presence of matter source
and to the equation for the NS 2-form field strength with the string
source
\begin{equation}\plabel{EEs}
R_{\underline{m}\underline{n}}-{1\over 2}R
g_{\underline{m}\underline{n}}
-{1\over 2}(\partial_{\underline m}\phi \partial_{\underline n}\phi
-{1\over 2}g_{\underline{m}\underline{n}}(\partial \phi)^2)
-{1\over
4}(H^{(3)}_{\underline{m}\underline{p}\underline{q}}H_{\underline
n}^{(3) \underline{p}\underline{q}}-{1\over
6}g_{\underline{m}\underline{n}}H^{(3) 2})
=e^{2\phi} T_{\underline{m}\underline{n}},
\end{equation}
\begin{equation}\plabel{Hs}
(-)^{(D-4)}\partial_{\underline
m}(e^{-2\phi}\sqrt{-g}H^{(3)\underline{m}\underline{n}\underline{p}})=
\int_{{\cal M}^D}\,d^D x \sqrt{-g}{\d^D (x-X(\xi))\over
\sqrt{-g}}
\int_{{\cal M}^2} d^2 \xi
\e^{ij}\partial_i X^{\underline n}\partial_j X^{\underline p}
\end{equation}
with $T_{\underline{m}\underline{n}}$ being the energy-momentum
tensor of the string
$$
T^{\underline{m}\underline{n}}=\int_{{\cal M}^D} {\d^D (x-X(\xi))\over
\sqrt{-g}}\int\, d^2 \xi \sqrt{-\g}\g^{ij}\partial_i X^{\underline m}
\partial_j X^{\underline n}.
$$

Now try to implement the consideration above to the case of the effective
action \p{Seff}. There are at least two different ways to establish the
T-duality covariance.

Following the first way, it is necessary to write down the source part of the
action \p{Seff} in the first order form by use of {\it worldvolume} field
$C_i$. Then, integrating out the $C$ field, one arrives at the T-duality
transformations \p{TRB}. In its turn, the bulk part of the action propagates
in the background defined by the line element
$$
d{\overline s}^2={\overline g}_{\underline{m}\underline{n}}dX^{\underline m}
\otimes dX^{\underline n}=
\hat{g}_{\underline{\tilde m}\underline{\tilde n}}
dX^{\underline{\tilde m}}\otimes dX^{\underline{\tilde n}}+g_{\underline{00}}
(dX^{\underline{0}}+A)\otimes (dX^{\underline{0}}+A).
$$ 
Therefore, we should
put it down in the KK manner, apply the rules \p{TRBKK} which is equivalent
to \p{TRB} together with the dilaton shift (cf. eq. \p{Tdil}) and lift it up.
This sequence leads to desired result (cf. eq. \p{Seffdual}).

The second way allows one to realize the gauged sigma-model-like technique at
the level of whole action \p{Seff} in uniform way having in mind that string
propagates in special background allowing for the isometry directions. To this
end, as in the sections before, extend the derivatives
to the covariant ones, i.e.
$$
dX^{\underline m} \longrightarrow
DX^{\underline m}=dX^{\underline m}+C k^{\underline m}
$$
with {\it target-space} additional field $C(X^{\underline{\tilde m}})$, 
and insert it into the action
\footnote{In \p{Seff1} $C_i$ is a pullback of the bulk one-form $C$, i.e.
$C_i=D_i X^{\underline m}C_{\underline m}\equiv \partial_i X^
{\underline{\tilde m}}C_{\underline{\tilde m}}$ since in the adapted 
coordinate frame $i_k C=0$.}
:
$$
S=\int_{{\cal M}^D}\, e^{-2\phi}\left[ {1\over (D-2)!}R^{{\underline
a}_1{\underline a}_2}(D)E^{{\underline a}_3}(D)\dots E^{{\underline
a}_D}(D)\e_{{\underline a}_1\dots {\underline a}_D}
+{1\over 2}D\phi \ast D\phi
+{1\over 2}H^{(3)}(D)\ast H^{(3)}(D) \right]
$$
$$
-{1\over 2}\int_{{\cal M}^D}\,d^D x \sqrt{-g}{\d^D (x-X(\xi))\over
\sqrt{-g}}~\times
$$
\begin{equation}\plabel{Seff1}
\int_{{\cal M}^2} d^2 \xi [
\sqrt{-\g}\g^{ij}D_i X^{\underline m}
D_j X^{\underline n}g_{\underline{mn}}(x)
+\e^{ij}D_i X^{\underline m}D_j X^{\underline n}
B^{(2)}_{\underline{n}\underline{m}}-2\e^{ij}C_i
\partial_j {\hat X}^{\underline 0}].
\end{equation}

The curl free field $C$ does not spoil the field content of the 
original theory. This is due to the symmetry under the local shifts in the
isometry direction which gives a possibility to fix either $C=0$ or
$X^{\underline 0}=0$ gauge. As in the case of the gauged sigma-model 
representation of string action \p{SD}, eq. \p{Seff1} reduces to the eq. 
\p{Seff}.

Consider now the first line of eq. \p{Seff1} and drop out for the moment the
curl free condition for $C$. Then, the background line element is defined by
$d{\overline s}^2=DX^{\underline m}\otimes DX^{\underline n}
{\overline g}_{\underline{m}\underline{n}}$ that is equivalent to the following
KK-type representation
\begin{equation}\plabel{KKint}
d{\overline s}^2={\hat g}_{\underline{\tilde m}\underline{\tilde n}}
dX^{\underline{\tilde m}}\otimes dX^{\underline{\tilde n}}+
g_{\underline{00}}(dX^{\underline 0}+C+A)\otimes
(dX^{\underline 0}+C+A).
\end{equation}
Therefore, the symbol ${\cal L}^{(D)}_{grav.+dil.+NS2.}(D)$ notifies that the
first line of \p{Seff1} is the compressed expression for theory
propagating in the KK-background \p{KKint}
and having therefore the form
\footnote{In \p{RKK} $H^{\prime (3)}=H^{(3)}+d(i_0 B^{(2)}) A$ and the Hodge 
star is the $(D-1)$-dimensional operator.}
$$
S=\int\, d^{(D-1)}x~ e^{-2\phi}\sqrt{g_{\underline{00}}}
\sqrt{-{g}}R
+\int_{{\cal M}^{(D-1)}}\, e^{-2\phi}\sqrt{g_{\underline{00}}}
[{1\over 2}d\phi \ast d\phi
+{1\over 2}d(\log |g_{\underline{00}}|) \ast d(\log |g_{\underline{00}}|)
$$
\begin{equation}\plabel{RKK}
+{1\over 2}g_{\underline{00}}
d(C+A)\ast d(C+A)
+{1\over 2}H^{\prime (3)}\ast H^{\prime (3)}+
{1\over 2}g^{-1}_{\underline{00}}d(i_0 B^{(2)})\ast d(i_0 B^{(2)})].
\end{equation}
Following our strategy it is necessary to find a solution to the equation of 
motion for the $C$ field and to insert it into the action. But as it has been 
pointed out in Section 2 the equation $C=dy$ still holds. Therefore, 
effectively, \p{RKK} does not
depend on $C$ and we return to the situation discussed previously in the
free case. After insertion of the solution for the $C$ equation of motion 
\p{A1}
and application of the KK T-duality rules, which, as previously mentioned,
have to be completed by the dilaton shift
\begin{equation}\plabel{Tdil}
\tilde{\phi}=\phi-{1\over 2}\log |g_{\underline{00}}|,
\end{equation}
the action \p{Seff1} converts into
$$
S=\int_{{\cal M}^D}\, e^{-2\tilde{\phi}}
\left[ {1\over (D-2)!}\tilde{R}^{{\underline
a}_1{\underline a}_2}E^{{\underline a}_3}\dots E^{{\underline
a}_D}\e_{{\underline a}_1\dots {\underline a}_D} 
+{1\over 2}d\tilde{\phi} \ast d\tilde{\phi}
+{1\over 2}{\tilde H}^{(3)}\ast {\tilde H}^{(3)}\right]
$$
\begin{equation}\plabel{Seffdual}
-{1\over 2}\int_{{\cal M}^D}\,d^D x \sqrt{-g}{\d^D (x-X(\xi))\over
\sqrt{-g}}
\int_{{\cal M}^2} d^2 \xi
[\sqrt{-\g}\g^{ij}\partial_i X^{\underline m}
\partial_j X^{\underline n}\tilde{g}_{\underline{mn}}(x)
+\e^{ij}\partial_i X^{\underline m}\partial_j X^{\underline n}
\tilde{B}^{(2)}_{\underline{n}\underline{m}}].
\end{equation}
Hence, T-duality indeed is the symmetry of the string effective action with
a string source.

\section{Discussion and conclusions}

Thus, we have demonstrated that T-duality still holds for the system comprising
the part of the bosonic sector of supergravity and the fundamental string 
source following two ways. The first one is the standard way and it shows the
influence of worldvolume duality to the target-space fields. The second way
emphasizes the role of isometry in T-duality consideration and gives the
possibility to realize in uniform manner the statement that both bulk and 
string parts of the
effective action with source propagate in background with isometry 
\footnote{Other aspects of space-time symmetry gauging in application e.g. to
the KK dimensional reduction can be found in \cite{nur}.}. However, following 
this way, we have to consider additional
target-space field rather than the worldvolume one. Due to this fact one has
also to take into consideration the bulk contribution to the equation of motion
for this new field. 
It looks surprisingly that establishing the invariance under the 
T-duality transformations sends in common to the uncoupling string case. 
At first glance one can expect to arrive to much more complicated problem
especially in the part of evaluation of the solution to the equation
of motion for the $C$ field deriving from the action \p{Seff1}. However, let's
analyse the problem from another point of view. As in the case of uncoupling
string, reviewed in Section 4, one can start from the consideration of the
action \p{Seffdual}. Then, inserting the extended derivatives and gauging out
the $\tilde{C}$ field, after application of T-duality rules we arrive at
the intermediate result, which is schematically the same as it was 
sketched in eq. \p{conn}
\footnote{This result remains the same for regarding $C$ as the bulk field
as well.}. Important point is that the KK gravity part in
the l.h.s. of this picture does not contain any dependence on 
$d\hat{X}^{\underline 0}$ being dual to the $\tilde{C}$ field. On the other
hand, in view of \p{conn}, the KK gravity part in the r.h.s. is also free from 
the dependence on $d\hat{X}^{\underline 0}$, but already with taking into 
account the solution to the equation of motion for the $C$ field. Therefore, we
conclude that no contribution comes from the KK gravity part to the equation of
motion for the $C$ field and therefore we can consider that $C=dy$ holds as
the ``Bianchi identity'' for the equation of motion for its dual.

Although our consideration is restricted to the case of one isometry 
direction, the proposed construction generalizes straightforwardly to the case 
of arbitrary number of commuting isometries. In view of gauging isometries
technique we may expect also that this scheme can be extended to the 
non-Abelian case. 

Another interesting problem is to apply this scheme to the case of open 
string and D-branes and to evaluate the T-duality power for the effective 
action in presence of the D-branes sources. Also it should be very attractive 
to realize completely supersymmetric picture of supergravity coupling with 
superstring as a source and to demonstrate the invariance under the (super) 
T-duality transformations. We hope to return to these problems in the
forthcoming publications.

\vspace{0.8cm}
{\bf \underline{Acknowledgements}}. We are grateful to Mario Tonin and 
Pier-Alberto Marchetti for interest to this work, and to Igor Bandos for
discussion and fruitful comments. We would like to thank Ian Jack and
Igor Rudychev for pointing out our attention on Refs. \cite{jjmo}, \cite{ps}, 
and all the colleagues for their encouragement.  
This work is supported in part by the
Ukrainian Ministry of Science and Education Grant N 2.5.1/52. Special
thanks to Nadezhda Sevritkina for encouragement and kind help.

\end{document}